# The efficiency of the likelihood ratio to choose between a t-distribution and a normal distribution

A decision must often be made between heavy-tailed and Gaussian errors for a regression or a time series model, and the t-distribution is frequently used when it is assumed that the errors are heavy-tailed distributed. The performance of the likelihood ratio to choose between the two distributions is investigated using entropy properties and a simulation study. The proportion of times or probability that the likelihood of the correct assumption will be bigger than the likelihood of the incorrect assumption is estimated.

J. Martin Van Zyl

*Department of Mathematical Statistics and Actuarial Science, University of the Free State, Bloemfontein 9300, South Africa.*

## 1. Introduction

A decision must often be made whether a regression or time series model has errors which are heavy-tailed distributed or not. The heavy-tailed t-distribution is frequently used and the choice is between this distribution and the alternative, normally distributed errors. The probability to make the correct decision when choosing between a model with normal errors and one with t-distributed errors will be estimated. The decision is based is based on the likelihood ratio which is the most powerful test for large samples.

It will be assumed that the errors or sample is a white noise series. It was found that the probability to make a wrong decision is very small in large samples and very high in small samples, and the error rate is a function of the degrees of freedom which is also the tail index of the t-distribution.

Let F denote a t-distribution (Student's t-distribution) and G a normal distribution, $x_1,...,x_n$, a random sample with distribution F. The interest is in the simple hypothesis

$$H_0 : x_j \sim F \text{ versus } H_1 : x_j \sim G \text{, for j=1,\ldots,n.}$$

The likelihood ratio is $\lambda_n = \prod_{j=1}^{n} (f(x_j)/g(x_j))$. It will be assumed that the variance of the t-distribution exist, that is the degrees of freedom is larger than 2, the variances of the two distributions are equal and the means equal to zero and known. The probability that $\lambda_n$ is less than one as a function of the sample size $n$ and degrees of freedom will be estimated using simulated data. That is the Type I error.

The expected value of the likelihood ratio is the relative entropy (Kullback-Leibler divergence), denoted by $D(F \| G)$, and this will be derived for the distributions and plotted as a function of the degrees of freedom. The relative entropy of two continuous distributions, *F* and *G*, is defined as

$$D(F \| G) = \int f(t) \log(f(t)/g(t)) dt \,. \qquad (1)$$

This can also be interpreted as a measure of "inefficiency" when using $G$ if the true distribution is $F$. $D(F \| G) \geq 0$, with equality if and only if $F = G$. The entropy of a distribution $F$ will be denoted $H(F)$ where

$$H(F) = -\int f(t)\log(f(t))dt. \tag{2}$$

It is shown in section 2 that $D(F \| G)$ is a minimum when the mean and variances of the two distributions are equal and the minimum value of $D(F \| G)$ is

$$D(F \| G) = \log\left(\frac{\Gamma((\nu+1)/2)}{\Gamma(\nu/2)(\nu/2)^{m/2}}\right) - \frac{\nu+1}{2}\left(\psi((\nu+1)/2) - \psi(\nu/2)\right) + 1/2 + (1/2)\log(\nu/(\nu-2)). \tag{3}$$

The lower bound on the Type I and II errors can be approximated when applying the Chernoff-Stein lemma (Chernoff, 1952), (Cover and Thomas, 1991). For a given Type II error rate, asymptotically the minimum Type II error denoted by $\beta$ when using the likelihood ratio to decide is

$$\beta = \exp(-nD(F \| G)). \tag{4}$$

The asymptotic lower of the Type I error is $\alpha = \exp(-nD(G \| F))$. In general $D(F \| G) \neq D(G \| F)$ and $D(G \| F)$ is a series with terms which are complicated integrals, and there is no simple closed form expression for $D(G \| F)$. In this work $D(G \| F)$ was calculated using numerical integration and was specifically used to estimate the asymptotic lower of the Type I error, $\hat{\alpha} = \exp(-nD(G \| F))$.

The Akaike Information Criterion (AIC), (Akaike, 1973) and Bayesian Information Criterion (BIC) (Schwarz 1976) have a penalty factor where the number of parameters is taken into account. AIC and BIC for a specific model can be written in the form $I = -2\log(L) + \varphi$ , where $L$ denotes the likelihood and a penalty term $\varphi$ which is a function of the number of parameters for model and sample size $n$. Using AIC model the t-distribution will be chosen if $\log(\lambda_n) + (p_1 - p_0) > 0$, where $p_0$ and $p_1$ denotes the number of parameters of the t- and normal distributions. If the mean is assumed to be known and equal to zero, the degrees of freedom and scale parameter must be estimated for the t-distribution and the MLE of the variance assuming a normal distribution, which means that the t-distribution will be penalized more than the normal. If the decision is based on the likelihood ratio, the error rates will be conservative compared to when the decision is made based on for example the AIC criterion.

It was found that a decision based on the likelihood ratio gives very small error rates in large samples and the heavier the tail of the t-distribution (or the smaller the degrees of freedom) the more efficient this method of choosing is. When the degrees of freedom is in the region of say eight and more, decisions with small error rates can only be made when the sample size is in the region of 300 and more.

## 2. The minimum relative entropy of the t- and normal distribution

The following expression for the entropy of the univariate standard t-distribution $F$ with $\nu$ degrees of freedom is given in this paper by Ebrahimi, Maascoumi and Soofi (1999):

$$H(F) = \log(\nu^{1/2} B(1/2, \nu/2) + \left(\frac{1+\nu}{2}\right)\left(\psi\left(\frac{1+\nu}{2}\right) - \psi\left(\frac{\nu}{2}\right)\right), \qquad (5)$$

and $\psi$ denotes the digamma function and $B$ the beta integral.

Consider the *m*-dimensional multivariate density t-distribution $F$ with $\nu$ degrees of freedom, parameters $\boldsymbol{\mu}^*, V$ and covariance matrix, $\Sigma^* = (\nu/(\nu-2))V^{-1}$ ,

$$f(\mathbf{x} \mid \boldsymbol{\mu}^*, V) = c(\nu + (\mathbf{x}-\boldsymbol{\mu}^*)' V (\mathbf{x}-\boldsymbol{\mu}^*))^{(m+\nu)/2},$$

$$c = \nu^{\nu/2} \Gamma((\nu+m)/2) \,|V|^{1/2} / \pi^{m/2} \Gamma(\nu/2),$$

and the entropy of this distribution (Guerrero-Cusumano (1996), is

$$H(F) = -\log\left(\frac{\Gamma((\nu+m)/2)}{\Gamma(\nu/2)(\nu\pi)^{m/2}}\right) + \frac{1}{2}\log(|V^{-1}|) + \frac{\nu+m}{2}\left(\psi((\nu+m)/2) - \psi(\nu/2)\right).$$

Let

$$g(\mathbf{x} \mid \boldsymbol{\mu}, \Sigma) = (2\pi)^{-m/2} |\Sigma|^{-1/2} \exp\left[-\frac{1}{2}(\mathbf{x}-\boldsymbol{\mu})' \Sigma^{-1} (\mathbf{x}-\boldsymbol{\mu})\right], \ \boldsymbol{\mu} : mx1, \Sigma > 0 : mxm,$$

denote a multivariate normal density. The entropy of G is

$$H(G) = (m/2)\log(2\pi e) + (1/2)\log(|\Sigma|).$$

The relative entropy between $F$ and a normal distribution is a minimum if the means and covariances are equal. A proof is given in appendix A. If the means and covariances are equal it can be shown that $D(F\,||\,G) = H(G) - H(F)$. Rao (1965) proved that of all *m*-dimensional distribution with covariance matrix Σ, the multivariate normal has the highest entropy. The relative entropy is:

$$D(F\,||\,G) = \log\left(\frac{\Gamma((\nu+m)/2)}{\Gamma(\nu/2)(\nu\pi)^{m/2}}\right) - \frac{1}{2}\log(|V^{-1}|) - \frac{\nu+m}{2}\left(\psi((\nu+m)/2\right) - \psi(\nu/2))$$
$$+(m/2)(\log(2\pi)+1) + (1/2)\log(|\Sigma|).$$

For $\Sigma = (\nu/(\nu-2))V^{-1}$:

$$D(F\,||\,G) = \log\left(\frac{\Gamma((\nu+m)/2)}{\Gamma(\nu/2)(\nu\pi)^{m/2}}\right) - \frac{\nu+m}{2}\left(\psi((\nu+m)/2\right) - \psi(\nu/2))$$
$$+(m/2)(\log(2\pi)+1) + (m/2)\log(\nu/(\nu-2)).$$

In the univariate case with $m=1$, this expression reduces to

$$D(F\,||\,G) = \log\left(\frac{\Gamma((\nu+1)/2)}{\Gamma(\nu/2)(\nu/2)^{1/2}}\right) - \frac{\nu+1}{2}\left(\psi((\nu+1)/2\right) \qquad (6)$$
$$-\psi(\nu/2)) + 1/2 + (1/2)\log(\nu/(\nu-2))$$

In figure 1 $D(F \| G)$ is plotted as a function of the degrees of freedom $\nu$.

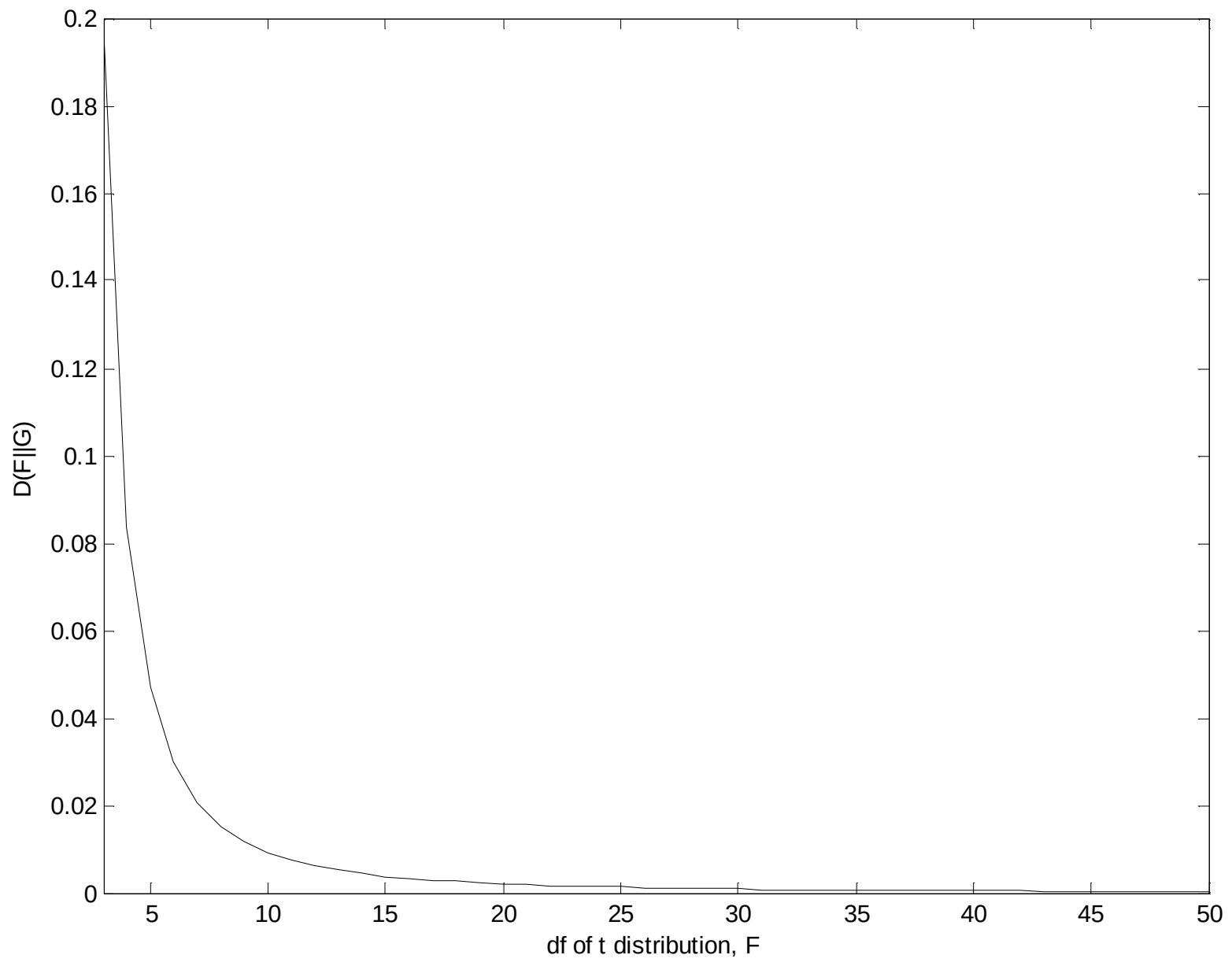

Figure 1. Plot of the relative entropy, D(G||F), between a standard t-distribution (F) and a standard normal distribution, as a function of the degrees of freedom of the t-distribution.

$D(G \| F)$ was calculated using numerical integration and was specifically used to estimate the asymptotic lower of the Type I error, say $\hat{\alpha} = \exp(-nD(G \| F))$. The approximate Type I error as a function of $n$ is plotted in figure 2, for $\nu = 4, 6, 8$.

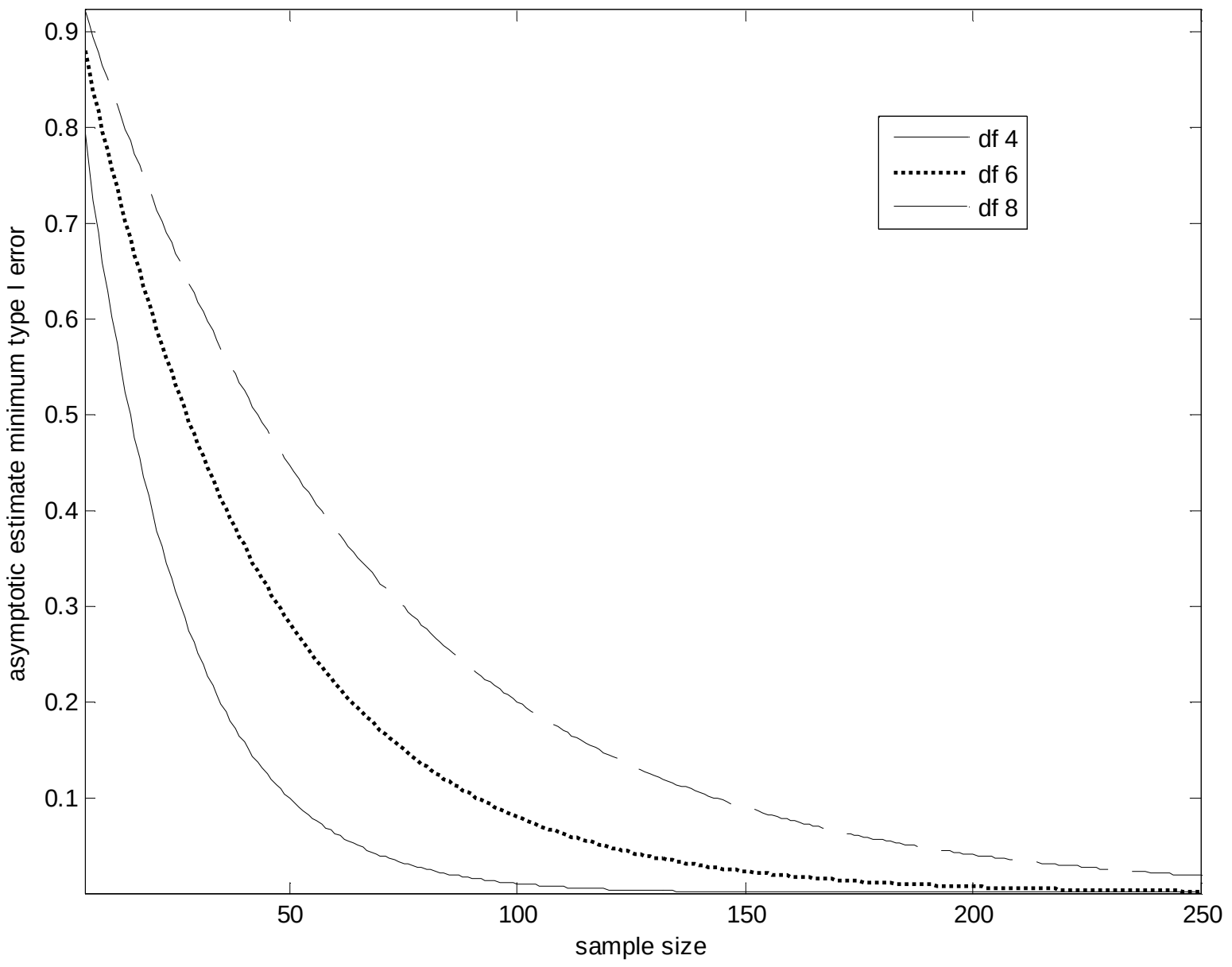

Figure 2. Asymptotic lower bounds of the Type I error when choosing between a normal and t-distribution, the data t -distributed.

## 3. Estimated error rates using simulated samples

At each sample of size n, k=1000 samples were simulated from a standard t-distribution with degrees of freedom $\nu$ and scale parameter $\sigma = 1$. It was assumed that the mean is known and equal to zero. The log-likelihood for a random sample, $x_1, ..., x_n$, calculated using the true known parameters, say $\boldsymbol{\theta} = (\sigma, \nu)'$ is

$$\log(L_n(F(\boldsymbol{\theta})) = n\log(\Gamma((\nu+1)/2)/\Gamma(\nu/2)(\nu\pi\sigma)^{1/2}) - ((\nu+1)/2)\sum_{j=1}^{n}\log(1+x_j^2/\nu\sigma).$$

The log-likelihood when normality is assumed and the variance estimated as $\hat{\sigma}^2$, using maximum likelihood is:

$$\log(L_n(G)) = -(n/2)\log(2\pi\hat{\sigma}^2) - (1/2)\sum_{j=1}^{n}(x_j^2/\hat{\sigma}^2) .$$

Let $\hat{\boldsymbol{\theta}}$ denotes the maximum likelihood estimator of the parameters of the t-distribution. In the simulation study a lower and upper bound for the error rates will be approximated by making use of large sample property

$$2[\log(L_n(F(\hat{\boldsymbol{\theta}})) - \log(L_n(F(\boldsymbol{\theta}))] \sim \chi_p^2 ,$$

where $p$ denotes the number of parameters estimated, which is 2 in this problem. Let $\alpha = 0.05$, and a 95% confidence interval for $L_n(F(\hat{\boldsymbol{\theta}}))$ is

$$L_n(F(\boldsymbol{\theta})) + (1/2)\chi^2_{2;\alpha/2} < L_n(F(\hat{\boldsymbol{\theta}})) < L_n(F(\boldsymbol{\theta})) + (1/2)\chi^2_{2;1-\alpha/2} . \qquad (7)$$

The maximum likelihood estimated degrees of freedom and scale parameter will not be calculated, but by using the above bounds, an approximate confidence interval for the maximum of the likelihood can be used. The error rates, that is when the log-likelihood for the normal is larger than the log-likelihood for the t-distribution will be calculated by using the three ratios: $[L_n(F(\boldsymbol{\theta})) + (1/2)\chi^2_{2;\alpha/2}]/L_n(G)$, $L_n(F(\hat{\boldsymbol{\theta}}))/L_n(G)$ and $[L_n(F(\boldsymbol{\theta})) + (1/2)\chi^2_{2;1-\alpha/2}]/L_n(G)$. The upper bound for the t-distribution will over estimate the error rate, and the lower bound will under estimate the error rates.

For each sample size, 1000 samples are generated and the proportion of time when normality would be accepted when it is a sample which is t-distributed plotted for $\nu = 4, 6, 8$ in figures 3, 4 and 5. The error rates decrease exponentially. It can be seen that the likelihood ratio performs weak in small samples, and this can be a more serious weakness than the bias and number of parameters when considering the performance of AIC and BIC. If for example AIC was used to make a decision, the error rates would be higher because the t-distribution has one more parameter than the normal.

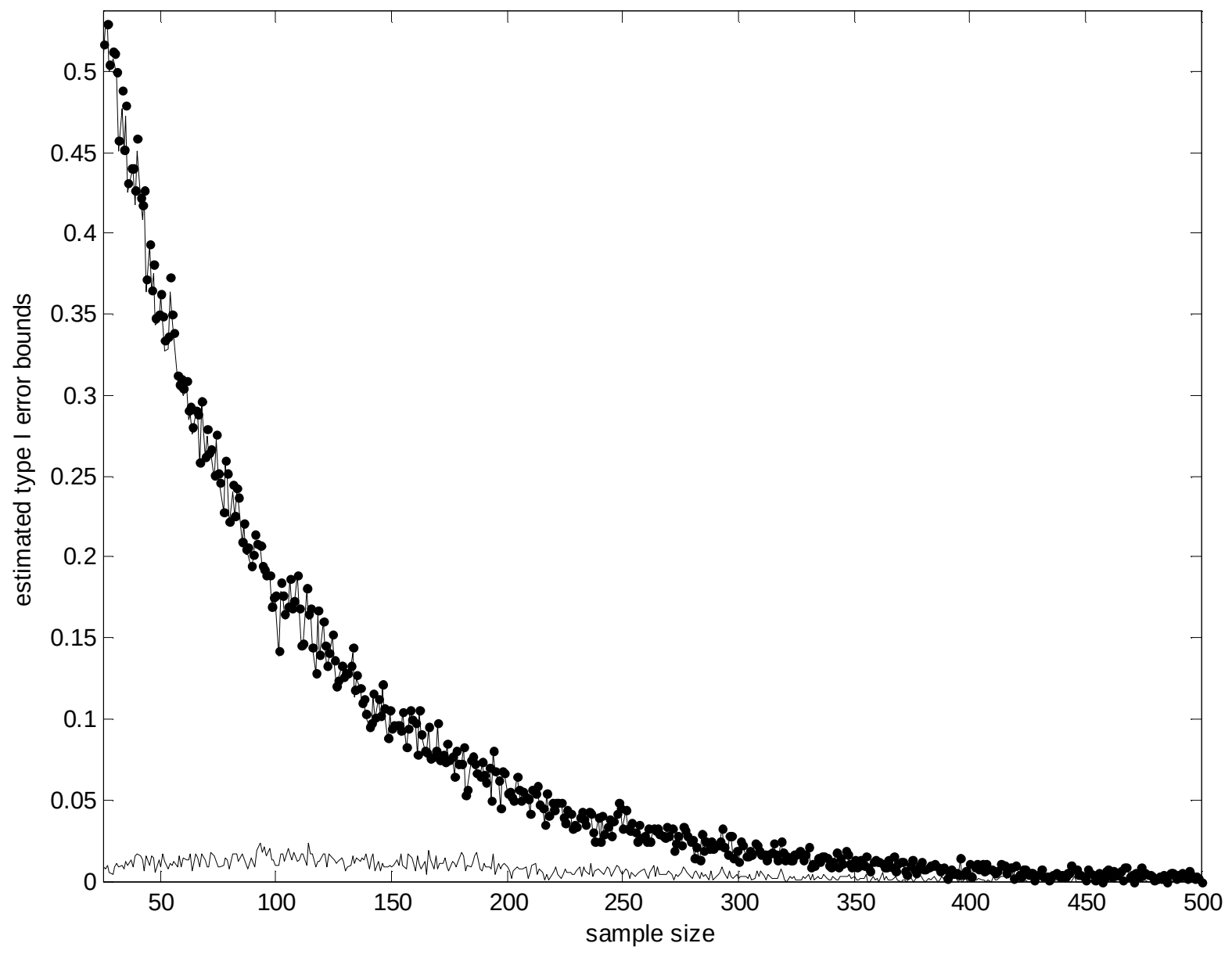

Figure 3. Lower and upper bounds for the Type I error, based on 1000 simulated samples for each sample size. The * denotes the likelihood ratio where the data is from a t-distribution and the likelihood ratio is calculated using the true parameters. The degrees of freedom of the t-distribution is 4.

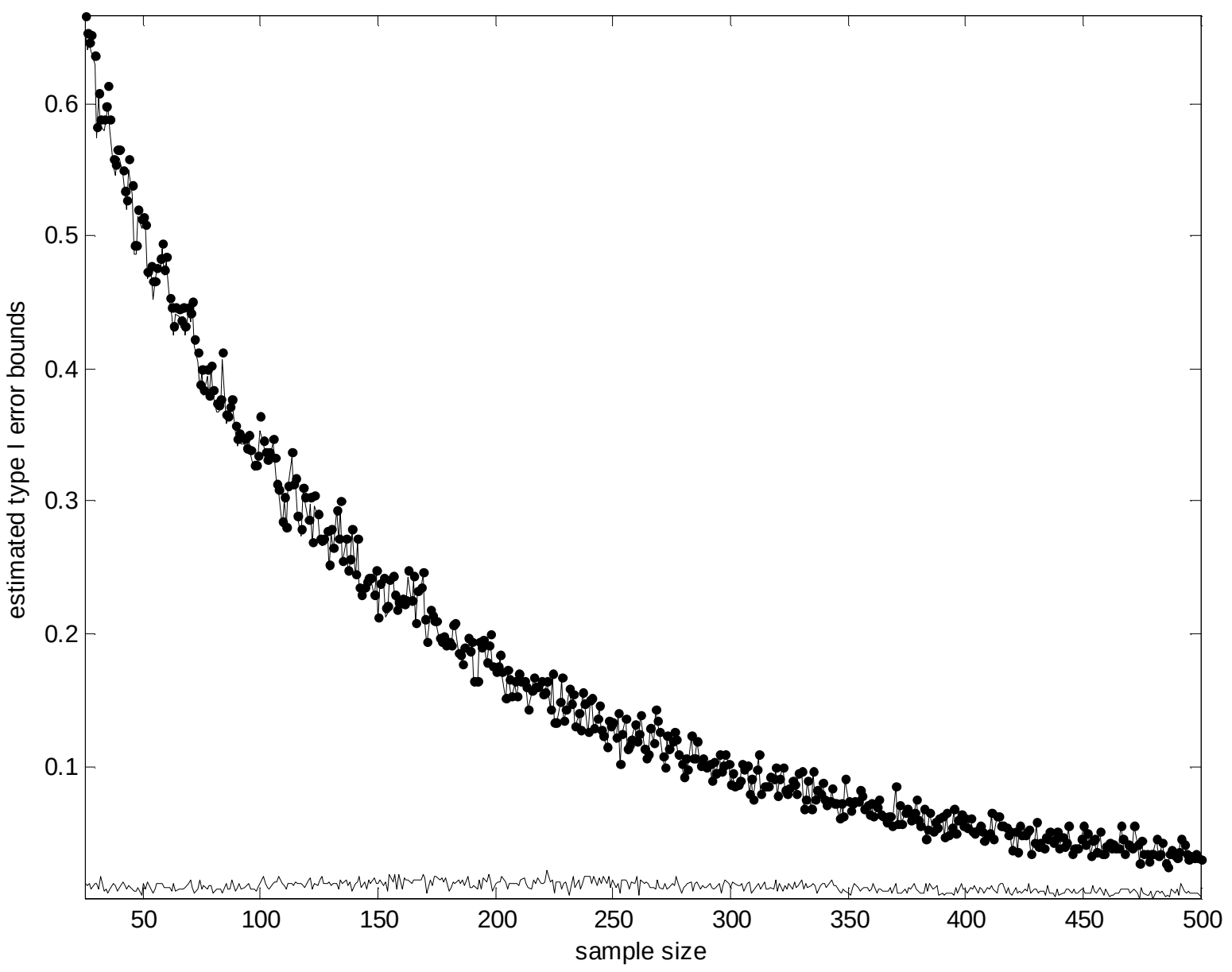

Figure 4. Lower and upper bounds for the Type I error, based on 1000 simulated samples for each sample size. The * denotes the likelihood ratio where the data is from a t-distribution and the likelihood ratio is calculated using the true parameters. The degrees of freedom of the t-distribution is 6.

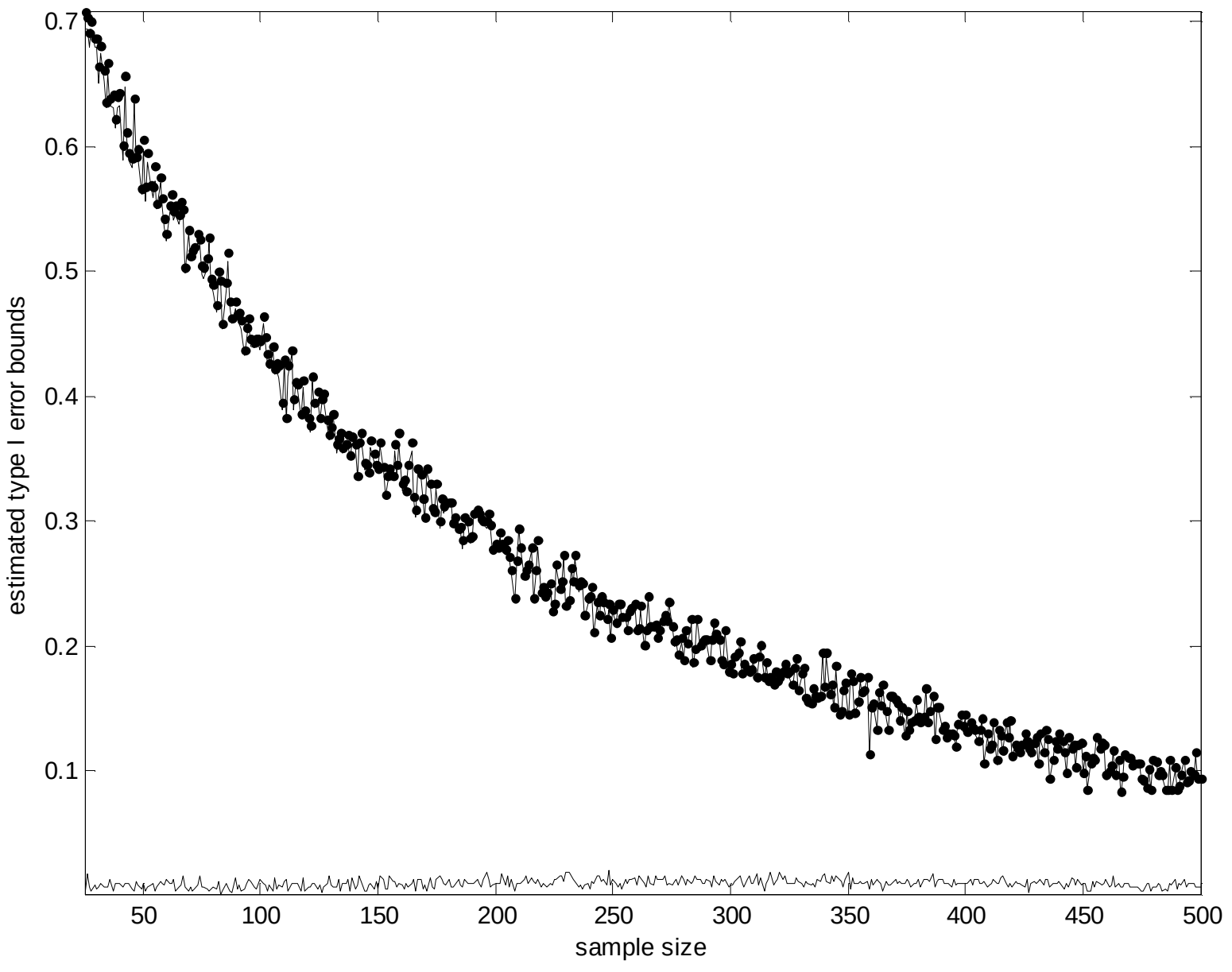

Figure 5. Lower and upper bounds for the Type I error, based on 1000 simulated samples for each sample size. The * denotes the likelihood ratio where the data is from a t-distribution and the likelihood ratio is calculated using the true parameters. The degrees of freedom of the t-distribution is 8.

## 4. Conclusions

Especially in financial time series, large sample sizes are available. It can be seen using the results in section 3, that very accurate decisions can be made when deciding on t distributed or normal errors for models. That is if more than $n = 500$ points are available and especially when the degrees of freedom is 6 or less. For samples of less than $n = 300$, the error rates are high, when using the likelihood ratio to decide. The degrees of freedom, $\nu = 5$, is often used in GARCH type models, assuring a finite $4^{th}$ moment and heavy tails. A minimum sample size of about $n = 250$ would assure accurate decisions.

In regression problems sample sizes are often much smaller than in time series and the likelihood ratio will be an acceptable procedure to decide between a model with normal errors and one with t distributed errors, only when the tails are very heavy, say for $\nu \leq 4$, for samples sizes less than in the region of $n = 100$. For lighter tails and small samples the use of the likelihood ratio is not more accurate than guessing.

## Appendix A

Let $P(\mathbf{x})$ denote a continuous multivariate distribution of dimension m, with the same support as the normal, finite second moments, mean $\boldsymbol{\mu}^*$ and covariance matrix $\Sigma^* > 0$. The relative entropy between $P$ and a normal distribution is a minimum if the means and covariances are equal. There are many variations of this result. Hernandez and Johnson (1980) showed that the parameter of the Box-Cox transformation which is optimal with respect to relative entropy, must be such that the first and second moments of the transformed variable and the normal to which it is transformed, are equal. Poland and Schachter (1993) applied this to estimate the parameters of mixtures of Gaussians.

Theorem 1: $P(\mathbf{x})$ denote a continuous multivariate distribution of dimension m, with

the same support as the normal, finite second moments, mean $\boldsymbol{\mu}^*$ and covariance matrix

$\Sigma^* > 0$. The relative entropy $D(P \| G)$ is a minimum if $\boldsymbol{\mu} = \boldsymbol{\mu}^*$ and $\Sigma = \Sigma^*$.

Proof: $D(P \| G) = \int_x p(x) \log(p(x)/g(x)) dx$

$$= -H(P) - \log((2\pi)^{-m/2} |\Sigma|^{-1/2}) + \frac{1}{2}\int_x p(\mathbf{x})(\mathbf{x}-\boldsymbol{\mu})'\Sigma^{-1}(\mathbf{x}-\boldsymbol{\mu})d\mathbf{x}.$$

Consider the expression:

$$\int_x p(\mathbf{x})(\mathbf{x}-\boldsymbol{\mu})'\Sigma^{-1}(\mathbf{x}-\boldsymbol{\mu})d\mathbf{x} = \int_x p(\mathbf{x})(\mathbf{x}-\boldsymbol{\mu}^* + \boldsymbol{\mu}^* - \boldsymbol{\mu})'\Sigma^{-1}(\mathbf{x}-\boldsymbol{\mu}^* + \boldsymbol{\mu}^* - \boldsymbol{\mu})d\mathbf{x}$$

$$= E_P(\sum_{j=1}^{m}\sigma^{jj}(x_j-\mu_j)^2 + 2\sum_{\substack{i\\ i\neq j}}\sum_{j}\sigma^{ij}(x_j-\mu_j)(x_i-\mu_i)) + (\boldsymbol{\mu}^*-\boldsymbol{\mu})'\Sigma^{-1}(\boldsymbol{\mu}^*-\boldsymbol{\mu})$$

$$= tr(\Sigma^*\Sigma^{-1}) + (\boldsymbol{\mu}^*-\boldsymbol{\mu})'\Sigma^{-1}(\boldsymbol{\mu}^*-\boldsymbol{\mu}).$$

The term $(\boldsymbol{\mu}^*-\boldsymbol{\mu})'\Sigma^{-1}(\boldsymbol{\mu}^*-\boldsymbol{\mu})$ is a minimum if $\boldsymbol{\mu} = \boldsymbol{\mu}^*$. Consider the term

$$\log(|\Sigma|) + tr(\Sigma^*\Sigma^{-1}) = -\log(|\Sigma^{*-1}\Sigma^*\Sigma^{-1}|) + tr(\Sigma^*\Sigma^{-1})$$

$$= -\log(|\Sigma^*\Sigma^{-1}|) + tr(\Sigma^*\Sigma^{-1}) + \log(|\Sigma^*|)$$

$$= \sum_{j=1}^{m}\lambda_j - \sum_{j=}^{m}\log(\lambda_j) + \log(|\Sigma^*|),$$

where $\lambda_j,\ j=1,...,m,$ are the characteristic roots of $\sum^{-1}\sum^*$. It can be shown that the expression is a minimum if $\Sigma = \Sigma^*$ and if $\Sigma^*\Sigma^{-1} = I_m$, the expression is equal to $\frac{1}{2}tr(I_m) + \frac{1}{2}\log(|\Sigma^*|) = \frac{m}{2} + \frac{1}{2}\log(|\Sigma^*|)$. The technique used to find the maximum without matrix differentiation is due to Watson (1964), Muirhead (1982, p85).

Thus, the relative entropy is a minimum for ***μ= μ****, Σ =Σ$^*$, and is equal to:

$$D(P\,||\,G) = -H(P) + (m/2)(\log(2\pi)+1) + (1/2)\log(|\Sigma|),$$

$$= H(G) - H(P),$$

where $H(P)$ denote the entropy of P.